\documentclass[aps, prd, letterpaper, 12pt, nofootinbib, superscriptaddress, longbibliography, notitlepage]{revtex4-1}
\usepackage[utf8]{inputenc}
\usepackage{amsmath,amssymb,amsfonts}
\usepackage{mathrsfs}
\usepackage{color}
\usepackage{graphicx} 
\usepackage[section]{placeins}
\usepackage[colorlinks=true,linkcolor=blue,citecolor=blue]{hyperref}
\allowdisplaybreaks

\pdfoutput=1 

\usepackage{graphicx}
\usepackage{hyperref}
\usepackage{aas_macros}
\usepackage{amsmath}
\usepackage{amssymb}
\usepackage{xcolor}
\usepackage{nccmath}
\usepackage{cleveref}

\def\beq{\begin{equation}\begin{aligned}}
\def\eeq{\end{aligned}\end{equation}}

\begin{document}

\title{\boldmath Dark Baryon Black Holes}

\author{Stefano Profumo,}
\email{profumo@ucsc.edu}
\affiliation{Department of Physics, 1156 High St., University of California Santa Cruz, Santa Cruz, CA 95064, USA}
\affiliation{Santa Cruz Institute for Particle Physics, 1156 High St., Santa Cruz, CA 95064, USA}

\begin{abstract}

\noindent We explore a novel mechanism for dark matter production through the formation of light black holes from the collapse of dark baryons in confining SU(N) gauge theories in the large $N$ limit. While glueballs and mesons cannot form black holes under physically reasonable conditions, we prove that for appropriate ranges of the confinement scale, quark masses,  number of colors $N$, and dark sector temperature, dark baryons can produce Planck-scale black hole relics in the early universe. 
Assuming the relics are stable, the abundance of both the dark baryon black hole population directly arising at confinement and that frozen in from dark glueball and meson pair annihilation are exponentially suppressed in $N$, leading to an upper limit $N\lesssim 100$ and of a few hundreds Planck units in mass for models where the black hole relics are the entirety of the dark matter. We present a detailed numerical study of the parameter space where this scenario is realized.
\end{abstract}

\maketitle
\newpage

\section{Introduction}

The nature of dark matter remains one of the most pressing mysteries in modern cosmology and particle physics. While numerous candidates have been proposed,  from weakly interacting massive particles to axions, the search for the fundamental nature of the dark matter and of the ``dark sector'' it resides in continues. In recent years, attention has also keenly turned to the possibility of dark sectors governed by gauge theories analogous to Quantum Chromodynamics (QCD) in the Standard Model.

The present study explores a novel mechanism for dark matter production in the early universe through the formation of black holes from dark baryons in confining SU(N) gauge theories. We demonstrate that under specific conditions, these dark baryon black holes (DBBHs) can form and, under suitable assumptions, potentially constitute the entirety of the observed dark matter.

Our approach builds upon the rich framework of dark SU(N) theories, which have been studied extensively for their potential to explain various astrophysical phenomena and to provide new dark matter candidates. By considering the large $N$ limit and the interplay between the confinement scale, quark masses, and dark sector temperature, we derive the physical conditions under which dark baryons can collapse into black holes, and compute the resulting black holes' mass and abundance.

The formation of primordial black holes (PBHs) as dark matter has been a subject of significant interest. However, our mechanism differs from traditional PBH formation scenarios in that it relies on the intrinsic properties of dark sector particles rather than density fluctuations in the early universe.

In this work, we first review the properties of hadrons in SU(N) gauge theories, focusing on their scaling behavior in the large $N$ limit. We then derive the conditions for dark baryon collapse into black holes and explore the cosmological implications of these objects. Finally, we present numerical results that demonstrate the viability of DBBHs as dark matter candidates and discuss the parameter space where this scenario can be realized.

\section{Dark confining SU(N) theories}

The question of the nature of dark matter and, more broadly, of physics beyond the Standard Model, has led to the investigation of dark, confining SU(N) and non-Abelian gauge theories. These theories offer a rich framework for addressing fundamental questions in particle physics and cosmology, providing novel approaches to dark matter candidates and mechanisms for explaining various astrophysical phenomena. This class of theories are particularly attractive due to their similarity to QCD in the Standard Model, and build on the rich  topical area of QCD's large $N$ limit studies. 
For instance, Ref.~\cite{Soni2017} investigated an SU(3) dark gauge theory coupled only to gravity. In their model, stable scalar glueballs emerge as potential dark matter candidates. 
Further extending this line of research, Ref.~\cite{Cai2020} examined an SU(4) gauge theory for composite dark matter. This construction builds upon previous studies of SU(N) theories, exploring the rich phenomenology that emerges from larger gauge groups. Ref.~\cite{Morrison:2020_1} explored both stable mesons and baryons, or a combination thereof, in single-flavor dark SU(N) (see also \cite{Fleming:2024flc, Mitridate:2017izz, Redi:2021kte, Asadi:2024bbq} for recent studies); a comprehensive review of similar models and theoretical constructions is presented in Ref.~\cite{Kribs:2016cew}.

Beyond their applications to dark matter, non-Abelian gauge theories have been explored for their potential to explain other cosmological phenomena. For instance, Ref.~\cite{Freese2016} proposed a dark sector model based on a non-Abelian gauge theory that links vacuum-dominated inflation with dark matter production. This approach offers a mechanism for smooth reheating and provides potential explanations for dark matter generation within a unified framework. 
The study of strongly coupled gauge theories has also yielded insights into physics beyond the Standard Model. These theories offer rich phenomenology that can be similar to or distinct from QCD, providing potential explanations for composite dark matter and shedding light on the nature of electroweak symmetry breaking~\cite{Ferretti2014}. 
%
%
Various other phenomenological predictions and implications include dark radiation~\cite{Buen-Abad2015},  sources for stochastic gravitational waves~\cite{Schwaller2015}, multi-component dark matter \cite{Morrison:2020yeg}, and dark sector self-interactions \cite{Tulin2018}.

In the present study we are especially concerned with determining whether high-multiplicity hadronic configurations can compress into regions sufficiently compact to reach the Schwarzschild radius, triggering gravitational collapse. Theoretical studies suggest that flux tubes connecting quarks during confinement can store sufficient energy density to drive the formation of PBHs, a mechanism previously studied for general confining gauge theories~\cite{Dvali:2021_1}. While many of these studies originate from visible-sector QCD analogies~\cite{Dvali:2011_1}, recent extensions to dark SU(N) scenarios have explored related concepts~\cite{Dvali:2022_1}.

In contrast to studies focused on black hole formation, much of the recent work on dark SU(N) gauge theories has centered on the dynamics of dense multi-quark hadrons formed during early-universe phase transitions~\cite{Asadi:2021_1,Gouttenoire:2023_1}. These studies describe scenarios in which heavy quarks are trapped in deconfined-phase pockets during first-order phase transitions, resulting in the formation of $N$-quark bound states either in bulk regions or at bubble wall boundaries. These dense hadronic configurations have been investigated primarily as potential dark matter candidates~\cite{Asadi:2021_2,Lonsdale:2017_1,Morrison:2020_1}, with analyses focusing on relic abundances, annihilation processes, and stability properties. However, direct connections between such compact, high-mass multi-quark bound states and gravitational collapse remain unexplored, to our knowledge. 
While progress has been made in understanding compactness thresholds~\cite{Dvali:2021_1,Dvali:2011_1}, and dense configurations in phase transitions~\cite{Asadi:2021_1,Gouttenoire:2023_1,Asadi:2021_2}, a comprehensive framework addressing the explicit gravitational triggers for black hole formation in dark-sector $N$-quark systems is still lacking. Our work seeks to build on these foundations by analyzing the critical mass, interaction effects, and confinement dynamics necessary for such phenomena.

\section{Hadron properties in the large $N$ limit}

In QCD, the SU(3) gauge theory describing the strong nuclear force, hadrons are composite particles formed by quarks bound by gluons~\cite{Gross:1973id}. The size of these hadrons is a result of the  balance between the confining force, characterized by the QCD scale $\Lambda_\text{QCD}$, and the properties of the constituent quarks, particularly their masses~\cite{Weinberg:1973ew}.
Extending this concept to a general SU(N) theory allows one to explore how hadron properties scale with the number of colors $N$, providing valuable insights into the large-$N$ limit of gauge theories~\cite{tHooft:1973alw}. In this broader context, we consider here a confinement scale $\Lambda$ analogous to $\Lambda_\text{QCD}$ in standard QCD, and a set of $N_f$ quark flavors all with the same mass $m_Q$, for simplicity.

The study of the large $N$ limit of SU(N) gauge theories, first proposed by 't Hooft~\cite{tHooft:1973alw}, has proven to be a powerful tool for understanding non-perturbative aspects of QCD and related strong interaction theories. This approach provides insight into the behavior of hadrons, including glueballs, mesons, and baryons, as the number of colors $N$ approaches infinity~\cite{Witten1979,Witten1979b}.
In this limit, the properties of such particles exhibit distinct scaling behaviors with respect to $N$, the confinement scale $\Lambda$, and the quark mass $m_Q$. Glueballs and mesons maintain finite masses as $N$ increases, while baryons become parametrically heavier~\cite{Manohar1998} -- a key feature in what follows. This scaling leads to a clear hierarchy in the hadron spectrum and, incidentally, offers explanations for various phenomena observed in QCD.

\subsection{Hadron Masses}

As $N$ approaches infinity, baryons, composed of $N$ quarks, exhibit a mass scaling that is linear with respect to $N$ 
This scaling relation emerges from several theoretical considerations. Witten \cite{Witten1979} first proposed this behavior arguing that in the large $N$ limit, baryons can be treated as a bound state of $N$ quarks in a mean-field approximation. The linear $N$-dependence arises from both the quark content and the binding energy of the system. 
The 't Hooft limit \cite{tHooft:1973alw}, where $\lambda = g^2N$ is held fixed as $N \to \infty$ (with $g$ the gauge coupling), provides a framework for understanding this scaling. In this limit, planar diagrams dominate, simplifying the many-body problem of $N$ quarks and leading to the observed linear scaling of baryon masses.
Lattice QCD simulations have provided numerical evidence supporting these large $N$ predictions, particularly in the pure gauge sector \cite{Lucini2004}. While direct experimental verification is challenging due to the fixed value of $N$=3 in our physical world, these theoretical insights have profound implications for our understanding of strong interactions and the structure of hadrons.


In the light quark limit, the masses of glueballs and mesons scale as $M \sim \mathcal{O}(1) \times \Lambda$, iwith $\Lambda$ the confinement scale, independent of $N$. For mesons with heavy quarks ($m_Q \gg \Lambda$), there is an additional contribution proportional to the quark mass, and the meson mass, in the heavy quark limit, goes as $2m_Q$; for intermediate $m_Q\sim\Lambda$, the mass $\sim\sqrt{m_Q\Lambda}$. Baryon masses, in contrast, grow linearly with N, scaling as $M_{\text{baryon}} \sim N \Lambda$ for light quarks and $M_{\text{baryon}} \sim N m_Q + \mathcal{O}(N \Lambda^2/m_Q)$ for heavy quarks~\cite{Jenkins1993}.\footnote{Note that with a single flavor, the lowest-lying bound state is the analogue of the Standard Model $\eta^\prime$, the single
pseudo-Nambu-Goldstone Boson meson associated with the breaking of the
accidental axial U(1) symmetry, with mass $\sim \Lambda/\sqrt{N}$ \cite{Veneziano:1976wm, Witten:1979kh,Morrison:2020yeg}.}

\subsection{Hadron size}

The size of the smallest hadron in large $N$ SU(N) is primarily determined by the two scales in the problem: the confinement scale $\Lambda$ and the quark mass $m_q$. In the heavy quark regime, the hadron size scales approximately as $R_H \sim 1/m_q$, reflecting the localization of the quark wave function due to its large mass~\cite{Isgur:1989vq}. This behavior is well-understood in the context of heavy quark effective theory~\cite{Neubert:1993mb}.
Conversely, for light quarks ($m_q \ll \Lambda$), the hadron size is governed by the confinement scale: $R_H \sim 1/\Lambda$~\cite{Shifman:1978bx}. This regime is particularly relevant for understanding the properties of pions and other light mesons in QCD, where chiral symmetry breaking plays a crucial role~\cite{Nambu:1960xd}.
Note that in the intermediate regime where $m_q \sim \Lambda$, both scales contribute significantly to the hadron size. A reasonable approximation in this case is $R_H \sim 1/\sqrt{m_q^2 + \Lambda^2}$, which smoothly interpolates between the heavy and light quark limits~\cite{Manohar:2000dt}.

The dependence of hadron size on the number of colors $N$ is generally weaker than its dependence on $\Lambda$ and $m_q$. As $N$ increases, the binding energy of the hadron typically increases, which could lead to a slight decrease in size~\cite{Witten:1979kh}. However, this effect is usually subdominant to the mass and confinement scale dependencies. For mesons and glueballs, we  estimate that the typical size of the smallest hadron of $N$ quarks in a confining SU(N) theory with coupling $\alpha_D\sim 1$ can be approximated as:

\begin{equation}
R_\text{hadron} \sim \frac{1}{\alpha_D\max(\Lambda, m_q)}
\end{equation}

The scaling with $N$ of {\em baryon} size is instead less trivial. Theoretical analyses and lattice QCD studies consistently demonstrate that while baryon masses scale linearly with $N$,  their spatial extent remains approximately constant. In the large-$N$ limit, baryons can be described as solitons or quantum Hall droplets \cite{Komargodski:2018omi}. The soliton picture, applicable for $N_f \geq 2$, arises from the topological properties of the chiral Lagrangian, with baryon number identified as a conserved topological charge. For $N_f = 1$, where Skyrmions are inapplicable, baryons can be interpreted as quantum Hall droplets with a size $L \sim \sqrt{N/T_{\text{sheet}}} \sim \mathcal{O}(1)$ in QCD units, where $T_{\text{sheet}} \sim N$ is the sheet tension \cite{Komargodski:2018omi}.

Lattice QCD studies provide numerical evidence for these scaling laws. Simulations with $N = 3, 5, 7$ demonstrate that baryon masses follow $M_N \propto N$, while their spatial extents remain consistent with $N$-independence \cite{DeGrand:2012hd}. For heavy quarks, in the mean-field approximation variational calculations applied to the baryon wavefunction which minimizes specific energy functionals  predict a baryon size $R_{\rm baryon} \sim \mathcal{O}(1)$, independent of $N$ \cite{Witten:1979kh}.
 For light quarks, instead, the analysis shifts to chiral Lagrangians with Skyrme terms. The soliton size is determined by balancing kinetic ($\sim \int (\partial U)^2$) and potential ($\sim \int (U + U^\dagger)$) terms, yielding:

\begin{equation}
R_{\rm baryon} \sim \frac{1}{\Lambda} \quad (\text{no } N\text{-dependence}).
\end{equation}
Higher-derivative terms, such as the Skyrme term $\sim (\partial U)^4$, stabilize the soliton without altering the $N$-scaling \cite{Adkins:1983ya}.

Lattice QCD studies provide numerical support for these theoretical predictions. Simulations across various $N$ values consistently show baryon masses scaling as $M_{\text{baryon}} \sim N$ while spatial extents remain $N$-independent \cite{DeGrand:2012hd}. These results, combined with effective field theory analyses, reinforce the robustness of the $\mathcal{O}(1)$ baryon size scaling in the large-$N$ limit.

The scaling behaviors for hadron masses and size in the large $N$ limit discussed above are summarized in Table \ref{tab:scaling}.
\begin{table}[t]
\centering
\caption{Scaling behavior of hadrons in the large $N$ limit}
\label{tab:scaling}
\begin{tabular}{|l|c|c|c|}
\hline
\textbf{Property} & \textbf{Glueballs} & \textbf{Mesons} & \textbf{Baryons} \\
\hline
Mass (light quarks) & $\mathcal{O}(1) \times \Lambda$ & $\mathcal{O}(1) \times \Lambda$ & $N \Lambda$ \\
Mass (heavy quarks) & - & $2m_Q + \mathcal{O}(\Lambda^2/m_Q)$ & $N m_Q + \mathcal{O}(N \Lambda^2/m_Q)$ \\
\hline
Size (light quarks) & $1/\Lambda$ & $1/\Lambda$ & $1/\Lambda$ \\
Size (heavy quarks) & - & $1/m_Q$ & $1/m_Q$ \\
\hline
\end{tabular}
\end{table}

\section{Dark Baryon Black Holes Formation}
The condition for hadron collapse into black holes is that the size of the hadron $R_H$ be smaller than the Schwarzschild radius corresponding to the hadron mass $M_H$,
\begin{equation}
    R_H<R_S(M_H)\sim\frac{M_H}{M_P^2},
\end{equation}
with $M_P$ the Planck mass. For glueballs, this condition directly implies $\Lambda>M_P$, which is nonphysical, as the confinement phase transition should occur after inflation and at sub-Planckian temperatures. For mesons, the same is true in the light-quark case, while in the heavy quark case one has
\begin{equation}
    \frac{1}{m_Q}<\frac{2m_Q}{M_P^2},
\end{equation}
implying a Planck-scale value for $m_Q$, which is arguably unnatural.

For baryons, instead, the condition reads
\begin{equation}
    \frac{1}{{\rm max}(\Lambda,m_Q)}<\frac{N{\rm max}(\Lambda,m_Q)}{M_P^2},
\end{equation}
thus, indicating with $X\equiv {\rm max}(\Lambda,m_Q)/M_P$,
\begin{equation}
    X>N^{-1/2},
\end{equation}
allowing for significantly sub-Planckian $\Lambda$ or $m_Q$ in the large $N$ limit.
Note that the black hole resulting from dark baryon collapse has a mass $NX$ in Planck units. Requiring that the object {\em is indeed a black hole} implies, in turn, that $M_{\rm BH}>M_P$, i.e. $NX>1$.

In summary, the condition for black hole formation are
\begin{equation}
    X>\frac{1}{N^{1/2}}>\frac{1}{N}.
\end{equation}

The time scale for dark baryon {\em collapse} into a black hole, corresponding to the hole's dynamical time scale, i.e. its light crossing time, reads
\begin{equation}
    t_d\sim \frac{N{\rm max}(\Lambda,m_Q)}{M_P^2}.
\end{equation}
Dark baryons therefore form black holes on a time scale that corresponds to the largest between the collapse and formation time scales (they need to coalesce before collapse), i.e. in a time
\begin{equation}
    t_{\rm formation}={\rm max}\left(\frac{N{\rm max}(\Lambda,m_Q)}{M_P^2},\ \frac{1}{\Lambda}\right),
\end{equation}
where we have assumed that the hadronization time-scale $t_{\rm baryon}\sim1/\Lambda$ is $N$-independent \cite{Komargodski:2018omi, Witten:1979kh, DeGrand:2012hd}.
If $\Lambda>m_Q$ this implies, since $\Lambda/M_P>N^{-1/2}$ for DBBH to form, that black holes form {\em after} dark baryon formation, i.e. $t_{\rm formation}=t_d>t_{\rm baryon}$. If instead $\Lambda<m_Q$ black holes may form promptly upon dark baryon collapse if 
\begin{equation}
    \frac{m_Q\Lambda}{M_P^2}<\frac{1}{N}.
\end{equation}

Dark baryons cannot form arbitrarily late, as the dark plasma abundance must disappear by Big Bang Nucleosynthesis (BBN), occurring at temperatures $T_{\rm BBN}\sim1$ MeV and at a time $t_{\rm BBN}\sim M_P/T_{\rm BBN}^2$. This implies the limit
\begin{equation}
    \frac{1}{\Lambda}<\frac{M_P}{T^2_{\rm BBN}}\ \Rightarrow \left(\frac{M_P}{T_{\rm BBN}}\right)\left(\frac{\Lambda}{M_P}\right)^{1/2}\sim 10^{22}X^{1/2}>1,
\end{equation}
which is easily satisfied in the region where DBBHs form.

A second constraint on $N$ stems from that fact that, in the large $N$ limit, the presence of a significant number of light particle species -- here the $N^2-1$ gluons associated with the dark SU(N) --  induces quantum corrections to Newton's constant $G$ through renormalization group effects. 
Experimental tests constrain deviations $\delta G/G$ at different scales, in particular 
    $\delta G/G \lesssim 10^{-2}$ at submillimeter scales \cite{Lee:2020zjt}. 
%
Calmet et al., Ref.~\cite{Calmet:2008tn}, derive the quadratic running
\begin{equation}\label{eq:RGMP}
    G^{-1}(\mu) \sim G^{-1} - \frac{N_{\rm dof}\mu^2}{12\pi},
\end{equation}
leading to an upper bound, given the constraint quoted above, of $N_{\rm dof}\sim N^2 \lesssim 10^{32}$ to prevent quantum gravity effects below $\mu \sim \text{TeV}$ (see also \cite{Ewasiuk:2024ctc} for a more extensive discussion on limits on a large number of dark degrees of freedom). This, in turn, implies that $N\lesssim 10^{16}$, a stronger limit than what quoted above from the collapse time-scale being shorter than the timescale of light element synthesis.

We note that the running in Eq.~(\ref{eq:RGMP}) affects the conditions above for (i) black hole collapse as well as for the requirements of (ii) the formed black hole to be trans-Planckian and (iii) the quark masses to be sub-Planckian. Given the speculative nature of a renormalization-group treatment of the running of the Planck scale, we postpone the detailed discussion of these effects to App.~\ref{app:RGMP} below\footnote{The Author is  grateful to Graham Kribbs for prompting him to explore the question of the running of the Planck scale}.

\section{Dark Baryon Black Holes Fate}


Upon confinement, dark glueballs and mesons are much more abundant than baryons, since, in the large $N$ limit, the dark baryon abundance is suppressed by a Boltzmann factor $\sim e^{-N}$. While it is possible that the lightest bound state be the dark matter (see e.g. \cite{Morrison:2020yeg}), as pointed out above stable dark glueballs or mesons do not lead to collapse to black holes. We will therefore assume  that either higher-dimensional operators mediate the decay of the lowest-mass bound state, or that the resulting relic abundance is small compared to the dark matter abundance. This ensures that the dark sector exclusively, or almost exclusively yields dark baryons, which we assume to be stable of very long-lived. We note that in confining dark sectors, the stability of the lightest baryon is often ensured by accidental symmetries, analogous to baryon number conservation in the Standard Model. In fact, the emergence of an accidental U(1) dark baryon number symmetry is a generic feature of confining theories with fermionic constituents, and the U(1) global symmetry arises naturally in the renormalizable Lagrangian of dark QCD-like theories \cite{Antipin:2015xia,Cline:2016nab}.

In addition to the population produced promptly at confinement, dark baryons $B$ are generically  ``frozen in'' by processes such as $(M,G)+(M,G)\to \bar B B$, with $M$ and $G$ the lowest-mass meson or glueball,  with a cross section $\sim e^{-2c_cN}$, with $c_c$ an ${\cal O}(1)$ factor describing the amplitude of the process and corresponding, in order of magnitude, to the absolute value of the logarithm of the probability to produce (annihilate) a single color quark pair \cite{Morrison:2020yeg} (the pair annihilation of dark baryons is negligible below the confinement phase transition, therefore the frozen in population is not depleted by subsequent dark baryon pair annihilation). 

We shall hereafter assume that the DBBHs formed upon dark baryon collapse do not undergo the semiclassical process of Hawking evaporation, and are therefore stable. This assumption is motivated by a number of considerations. First, when $M_{\rm DBBH} \sim M_P$, the Schwarzschild radius becomes comparable to the Planck length, rendering spacetime fluctuations non-negligible \cite{Dvali:2020wft}. 
The derivation of Hawking radiation assumes field modes with arbitrarily high frequencies near the horizon \cite{Helfer:2000fg}. Close to the Planck scale, these modes probe sub-Planckian distances where: (1) The semiclassical approximation breaks down, (2) Quantum gravity effects dominate, and (3) The black hole's entropy $S = k_B A/4\ell_P^2$, with $\ell_P$ the Planck length, approaches $\mathcal{O}(1)$, suggesting information capacity collapse \cite{Page:2004xp}. Recent analyses show suppressed emission rates $\Gamma \sim e^{-S}$ for $M < q m_P$ ($q \sim 0.5$) due to memory burden effects \cite{Auffinger:2022khh}. 
Lattice simulations in analogue gravity models demonstrate evaporation halting when $R_s \approx 2\ell_P$, consistent with energy conservation in modified dispersion relations \cite{Corley:1998rk}. The loss of thermodynamic description at $T_H \sim T_P/8\pi$ further suggests kinematic closure of evaporation \cite{Barcelo:2005fc}. We consider the stability of DBBHs, therefore, as a rather plausible possibility. Should that possibility be violated, DBBHs would evaporate very quickly, but possibly produce, upon evaporation, cosmologically interesting relics, including, potentially, the dark matter \cite{Morrison:2018xla}.

The comoving number density $Y=n/s$ of dark baryons  produced at confinement is $Y_\Lambda\sim e^{-2cN}$, where the exponential suppression, again, stems from the probability of quarks of different colors ``finding'' one another. Dark baryons are additionally ``frozen in'' from baryon-antibaryon production from meson or glueball collisions, as explained above. The  evolution of the frozen-in dark baryon comoving number density follows the Boltzmann equation\footnote{Note, again, that we neglect the pair-annihilation of dark baryons, which is Boltzmann-suppressed compared to the abundance of dark mesons and glueballs.}
\begin{equation}
    \frac{dY}{dx}\sim  \frac{s_{\rm SM}(x)\langle \sigma_{(M,G)+(M,G)\to \bar B B} v \rangle}{x\ H(x)}Y^2_{{\rm eq},(M,G)}
\end{equation}
from the confinement temperature down to lower temperatures (as we will see, the final temperature is irrelevant), where as usual $x=m_{G,M}/T_{\rm SM}\sim \Lambda/T_{\rm SM}$, $$s_{\rm SM}(x)\sim g_{*{\rm SM}}\ T_{\rm SM}^3=g_{*{\rm SM}}\left(\frac{\Lambda}{x}\right)^3,$$
with $g_{*,{\rm SM}}$ the number of Standard Model effective relativistic degrees of freedom at high temperature,
$$\langle \sigma_{(M,G)+(M,G)\to \bar B B} v \rangle\sim \frac{e^{-2c_c N}}{\Lambda^2},$$ with $c_c$ parameterizing the collision term \cite{Morrison:2020yeg}, with the assumption of negligible velocity dependence, and the Hubble rate $$H(x)\sim g_{*{\rm SM}}^{1/2}\frac{\Lambda^2}{{M_P\  x^2}}.$$ The equilibrium comoving number density of mesons/glueballs is initially, around the confinement scale, relativistic, thus
\begin{equation}
    Y_{{\rm eq},(M,G)}
    \sim \frac{N^2}{g_{*{\rm SM}}}\xi^3,
\end{equation}
with $\xi$ the dark-to-visible sector temperature ratio.

The Boltzmann equation then reads, recalling that the pair annihilation of dark baryons is negligible because of Boltzmann suppression,
\begin{equation}
    \frac{dY}{dx}\sim e^{-2c_c N}\left(\frac{M_P}{\Lambda}\right)\frac{N^4\xi^6}{g^{3/2}_{*{\rm SM}}}\frac{1}{x^{2}},
\end{equation}
and its solution is, approximately, 
\begin{equation}
    Y_{\rm DBBH}\sim e^{-2c_c N}\left(\frac{M_P}{\Lambda}\right)\frac{N^4\xi^6}{g^{3/2}_{*{\rm SM}}}{\xi}\sim \left(N^4 e^{-c_* N}\right)\frac{\xi^7}{g^{3/2}_{*{\rm SM}}}\left(\frac{M_P}{\Lambda}\right),
\end{equation}
where we absorbed the initial population produced at confinement by redefining \cite{Morrison:2020yeg}
$$c_*=-2(c_c+c)$$  and we neglected both the low-temperature component and finite-density effects\footnote{I am thankful to Tim Cohen for suggesting to look into this point.}, as usual.

The abundance of dark baryons collapsing to DBBHs today is then given by
\begin{equation}
    \rho_{\rm DBBH}=M_{\rm DBBH}n_{\rm DBBH}=N{\rm max}\left(\Lambda,m_Q\right)Y_{\rm DBBH}s_0={\rm max}\left(\Lambda,m_Q\right)s_0\left(N^5 e^{-c_* N}\right)\frac{\xi^7}{g^{3/2}_{*{\rm SM}}}\left(\frac{M_P}{\Lambda}\right),
\end{equation}
with $s_0$ the entropy density today.
The DBBH abundance relative to the dark matter abundance is thus
\begin{equation}
    \Omega_{{\rm DBBH},\Lambda}=\frac{\rho_{\rm DBBH}}{\rho_{\rm DM}}\sim\frac{s_0\ M_P}{\rho_{\rm DM}}\left(N^5 e^{-c_* N}\right)\frac{\xi^7}{g^{3/2}_{*{\rm SM}}}
\end{equation}
for $m_Q<\Lambda$ and
\begin{equation}
    \Omega_{{\rm DBBH},m_Q}\sim\frac{s_0\ M_P}{\rho_{\rm DM}}\left(N^5 e^{-c_* N}\right)\frac{\xi^7}{g^{3/2}_{*{\rm SM}}}\left(\frac{m_Q}{\Lambda}\right)
\end{equation}
for $m_Q>\Lambda$. Note that from the constraint that the dark plasma disappears by BBN (see above),
\begin{equation}
    \frac{m_Q}{\Lambda}\ll \frac{1}{N}\left(\frac{M_P}{T_{\rm BBN}}\right)^2\sim \frac{10^{44}}{N},
\end{equation}
and, demanding that $m_Q<M_P$ and $\Lambda\gtrsim 10$ TeV, 
\begin{equation}
    \frac{m_Q}{\Lambda}\ll 10^{15}.
\end{equation}
In sum, the constraint is:
\begin{equation}
    \frac{m_Q}{\Lambda}\ll {\rm min}\left(\frac{10^{44}}{N},10^{15}\right).
\end{equation}
Inserting numerical values for the constants above, we find that 
\begin{equation}\label{eq:lambda}
\Omega_{{\rm DBBH},\Lambda}\sim (5\times 10^{24})\ \xi^7  N^5 e^{-c_* N},
\end{equation}
\begin{equation}\label{eq:mQ}
\Omega_{{\rm DBBH},m_Q}\sim (5\times 10^{24})\ \xi^7  N^5 e^{-c_* N}\left(\frac{m_Q}{\Lambda}\right)\ll  (5\times 10^{24})\ \xi^7  N^5 e^{-c_* N}\ {\rm min}\left(\frac{10^{44}}{N},10^{15}\right).
\end{equation}

Two remarks are  in order: First, note that 
we did not entertain an asymmetry in the dark baryonic sector, that would linearly suppress the results in Eq.~(\ref{eq:lambda}-\ref{eq:mQ}) above; secondly, while there exist constraints on the ratio of dark-to-visible dark sector temperatures (see e.g. \cite{Baumann:2016wac,Garcia:2020eof,Amin:2022pzv}), they do not apply below the confinement phase transition, which, here, happens at very early times and well before BBN.
\begin{figure}
    \centering
    \mbox{\includegraphics[width=0.5\linewidth]{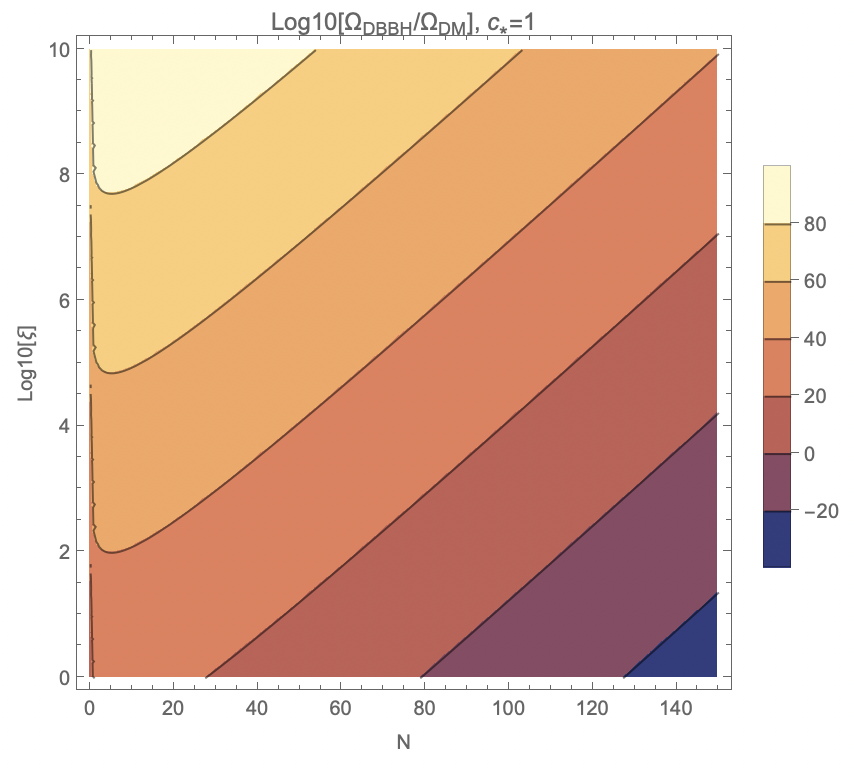}\quad \includegraphics[width=0.5\linewidth]{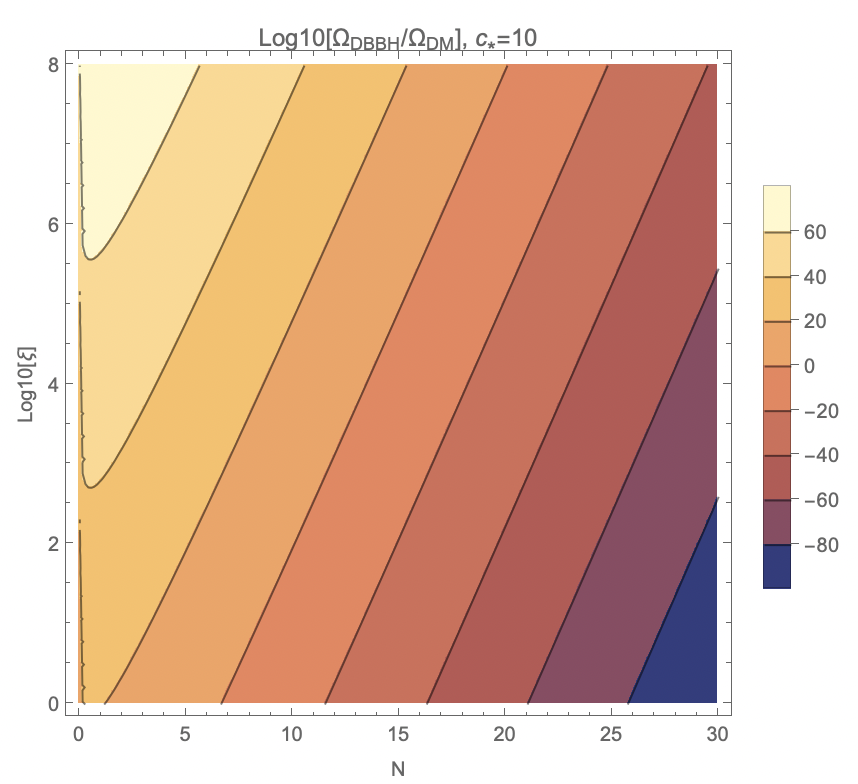}}
         \caption{Contours of relic DBBH relic abundance in units of the observed cosmological dark matter abundance, $\Omega_{\rm DBBH}/\Omega_{\rm DM}$, on the ($N,X$) plane, on a linear and Log10 scale, respectively, for $c_*=1$ (left) and 10 (right); the line labeled with ``0'' corresponds to a relic density equal to the cosmological dark matter density.}
    \label{fig:Omega}
\end{figure}
\section{Numerical Results}

\begin{figure}
    \centering
    \mbox{\includegraphics[width=0.47\linewidth]{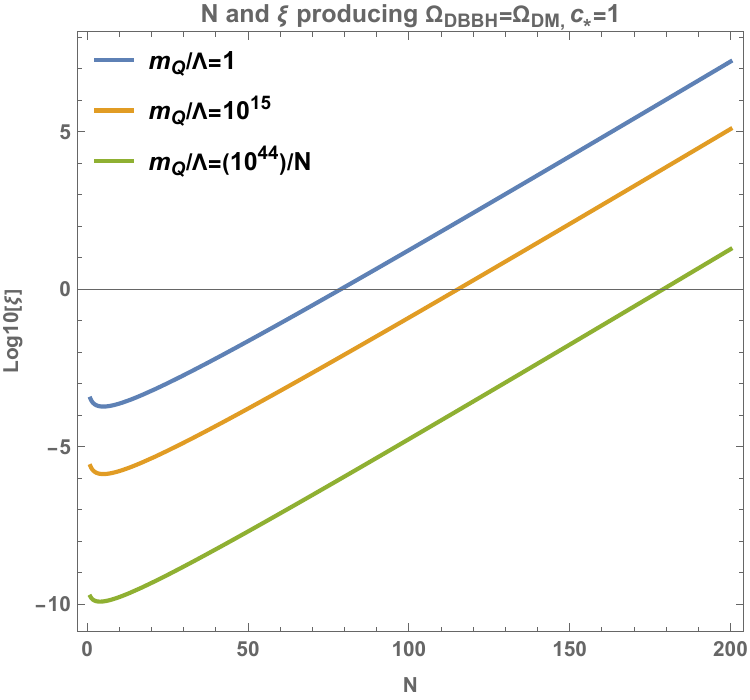}\quad \includegraphics[width=0.47\linewidth]{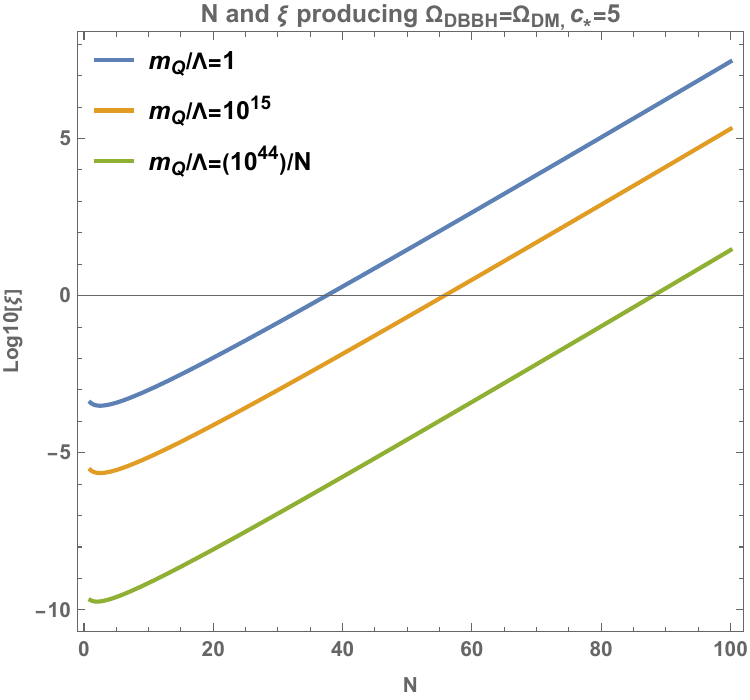}}
         \caption{The dark-to-SM temperature ratio $\xi$ necessary, as a function of $N$, for $m_Q>\Lambda$, to produce a DBBH relic abundance matching the observed dark matter abundance, for $c_*=1$ (left) and 5 (right). The horizontal ``0'' line corresponds to identical dark and visible sector temperatures.}
    \label{fig:mQ}
\end{figure}

Fig.~\ref{fig:Omega} shows the relic abundance of DBBHs  in units of the observed cosmological dark matter abundance, $\Omega_{\rm DBBH}/\Omega_{\rm DM}$, on a Log10 scale, on the plane defined by $N$ and the dark-to-visible sectors temperature ratio $\xi$, the latter on a Log10 scale, for $c_*=1$ (left) and 10 (right). Notice the different color scales in the left and right panels. The areas in the lower, right corners, below the ``0'' contour, produce an under-abundant density of DBBH for them to be the dark matter. This is, of course, not ruled out, but DBBHs are not the majority of the dark matter in that part of the parameter space. Notice also that, in the opposite upper-left corner of very hot dark sectors and relatively low $N$, the abundance can be so large that it could survive, in principle, the dilution factor from an inflationary period consisting of $N_e$ e-foldings, if 
\begin{equation}\frac{\Omega_{\rm DBBH}}{\Omega_{\rm DM}}\times e^{-3N_e}\sim1.
\end{equation}.
Fig.~\ref{fig:mQ} shows the values of $\xi$ necessary, as a function of $N$, for $m_Q>\Lambda$, to produce a DBBH relic abundance matching the observed dark matter abundance, for $c_*=1$ (left) and $c_*=5$ (right). The horizontal ``0'' line corresponds to identical dark and visible sector temperatures. The blue lines assume $m_Q/\Lambda=1$, while the orange and green lines saturate the limits on the ratio $m_Q/\Lambda$ from precision gravity constraints and from BBN. 
The plots illustrate how for low-enough $c_*$ and hot dark sectors, despite the exponential suppression, $N$ can exceed 100, resulting in DBBH possibly over two orders of magnitude larger than the Planck scale (see fig.~\ref{fig:LambdaMass} below).

The left panel of fig.~\ref{fig:LambdaMass} shows the same as in fig.~\ref{fig:mQ}, but for $\Lambda>m_Q$ and for various values of $c_*$; the figure illustrates how, for large values of $c_*$, the correct relic abundance can only be achieved for  low $N$ (resulting in light DBBH close to the Planck scale) and very hot dark sectors. For lower values of $c_*$, instead, $N$ can approach, again, values around 100.

The right panel of fig.~\ref{fig:LambdaMass} shows, on the ($X,N$) plane, where, as a reminder, $$X\equiv{\rm max}\left(\Lambda, m_Q\right)/M_P,$$on a Log10 scale, contours of the DBBH mass in grams (thus the ``-3'' line corresponds to DBBH of one milligram etc.). The figure also shows the line, in yellow, corresponding to $X=N^{-1/2}$, to the right of which collapse into DBBH occurs; the region to the left of the yellow line does not form black holes instead. Sub-Planckian black holes correspond in the region to the top-right of the green line. Given that $N\lesssim 100$, the figure implies that DBBH have, at most, a mass $N{\rm max}(\Lambda,m_Q)\sim 100\ M_P$. 
\begin{figure}
    \centering
    \mbox{\includegraphics[width=0.47\linewidth]{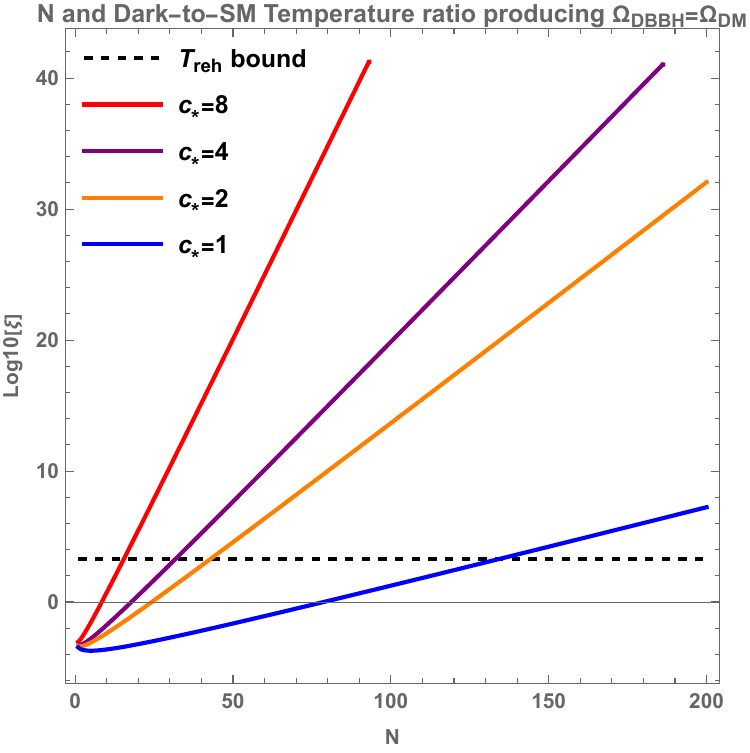}\quad \includegraphics[width=0.47\linewidth]{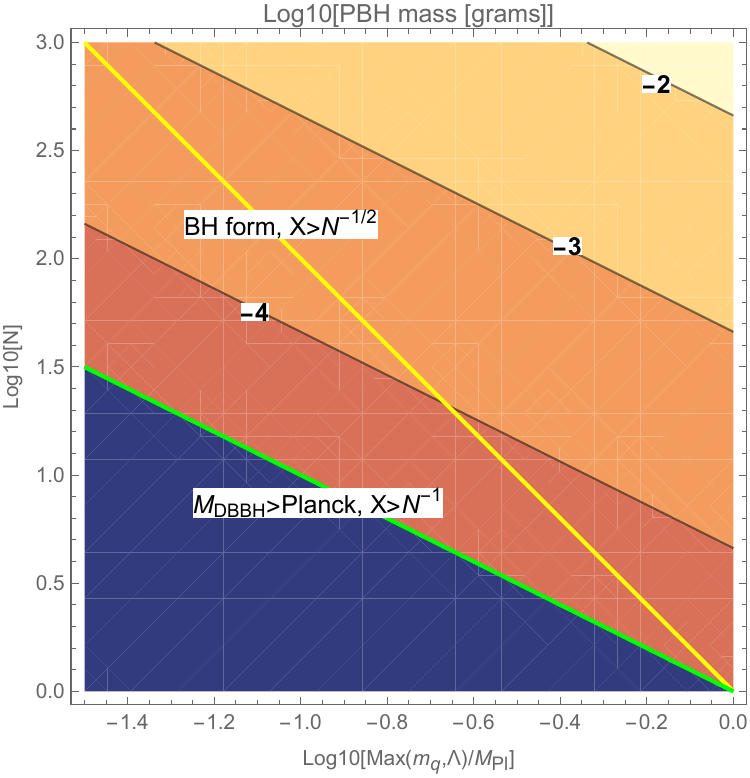}}
         \caption{Left: the dark-to-SM temperature ratio $\xi$ necessary, as a function of $N$, for $m_Q<\Lambda$, to produce a DBBH relic abundance matching the observed dark matter abundance, for various values of $c_*$; Right: contours of the relic DBBH's mass in grams, on a Log10 scale, on the $(X,N)$ plane. }
    \label{fig:LambdaMass}
\end{figure}

A final remark pertains to the scale of the confinement phase transition. The latter must occur at temperatures below the reheating temperature, $\Lambda<\xi T_{\rm reh}$, least the abundance of the produced DBBH be exponential diluted by the inflationary epoch\footnote{Again, it may well be that the DBBH abundance be so large that it would survive inflationary dilution, see above.}. It is important, therefore, to summarize the state of the art for inflationary models with the highest-possible reheating temperature.

Analytical studies frequently identify $T_{\text{reh}} \sim 10^{15} \, \text{GeV}$ as the upper bound, derived from the inflaton's energy density at the end of inflation and slow-roll conditions limiting the scale of inflation \cite{Lahanas:2022}. Observational constraints, such as measurements of the scalar spectral index $n_s$ and non-detection of primordial gravitational waves, further restrict $T_{\text{reh}}$, with stricter limits in scenarios with fewer e-folds during inflation, where $T_{\text{reh}} \sim 10^{13} \, \text{GeV}$ has been proposed \cite{Lorshbough:2015}. 
We also note that thermodynamical processes such as preheating through parametric resonance can create transient maximum temperatures $T_{\text{max}} > T_{\text{reh}}$, although backreaction effects and thermalization bottlenecks prevent $T_{\text{reh}}$ itself from approaching trans-Planckian scales \cite{Shtanov:1994ce,Haque:2020}. 

Reheating temperatures approaching the Planck scale, however, are not in principle firmly ruled out.  First, modified inflationary scenarios like 
$k$-inflation \cite{ArmendarizPicon:1999rj} alter the relation between 
$r$ and the energy scale, introducing a reduced sound speed $C_s$  for perturbations $$r=-8C_sn_t\quad (C_s<1),$$ relaxing the bound on the inflationary energy scale. Similarly, multifield models \cite{Wands:2007bd} modify the consistency relation through adiabatic mode evolution, weakening the 
$r$-energy scale linkage. Second, a speculative resolution posits that quantum gravity effects might suppress tensor mode production entirely if gravity remains non-quantized \cite{Barrau:2019cuo}, decoupling $r$ from the inflationary energy scale. While these approaches circumvent observational limits, achieving scalar spectrum normalization at such high energies remains challenging, necessitating careful model-building to reconcile the required reheating temperature with cosmological consistency conditions.

In the present context, however, one should not compare the dark sector temperature with limits on $T_{\rm reh}$ but, rather, the Standard Model temperature, $T_{\rm SM}=T_{\rm DS}/\xi$. Therefore, since for $\Lambda>m_Q$ we find, from fig.~\ref{fig:LambdaMass} that $\Lambda\gtrsim 0.2 M_P$, the condition $\Lambda<\xi T_{\rm reh}\lesssim\xi\ 10^{15}$ GeV implies that $\xi\gtrsim 2\times 10^3$. We indicate this constraint with a horizontal dashed line in fig.~\ref{fig:LambdaMass}. The $\xi\lesssim 2\times 10^3$ is not firmly ruled out because of the considerations above. 
For $m_Q>\Lambda$, instead, $\Lambda$ can be much below $m_Q\gtrsim 0.2 M_P$ (again from fig.~\ref{fig:LambdaMass}) and the dark sector need not be as hot, meaning that any value of $\xi$ is possible.

A few remarks are additionally in order here: first, accretion onto Planck-scale relics of mass $M$ is always irrelevant. To prove this, we compare the mass  accretion rate with the Hubble rate at temperature $T$,
\begin{equation}
    \frac{1}{M}\frac{dM}{dt}\sim \frac{1}{M}(\sigma \cdot \rho(T))\sim \frac{1}{M}\left(\frac{M}{M_P^2}\right)^2T^4;\qquad H\sim \frac{T^2}{M_P},
\end{equation}
which shows that 
\begin{equation}
    \frac{1}{H}\left(\frac{1}{M}\frac{dM}{dt}\right)\sim \frac{M_P}{T^2} \frac{1}{M}\left(\frac{M}{M_P^2}\right)^2T^4\sim\left(\frac{M}{M_P}\right)\left(\frac{T}{M_P}\right)^2\ll1,
\end{equation}
since the first factor in the last equation is at most $\sim100$, while the latter is $\lesssim 10^{-6}$ given limits on the maximal reheating temperature.

Secondly, the merger rate for Planck-scale relics is also negligible compared to the Hubble rate. Peters' formula for binary formation from gravitational radiation \cite{Peters:1963} indicates that the merger time scale for black holes of mass $M$ is approximately
\begin{equation}
    \tau_{\rm merger}\sim M_P^6\frac{a^4}{M^3},
\end{equation}
with $a$ the orbital separation, given by the $n^{-1/3}$, with $n$ the number density of black holes,
\begin{equation}
n(T)\lesssim \frac{s(T)}{s_0}\frac{\rho_{\rm DM}}{M}\sim g_{*,{\rm SM}}\ T^3\ \frac{10^{-4}\ {\rm GeV}}{M}\sim 10^{-21}\ T^3\ \frac{M_P}{M}
\end{equation}
thus
\begin{equation}
    a\sim n^{-1/3}\sim 10^{7}\frac{1}{T}\left(\frac{M_P}{M}\right)^{-1/3}.
\end{equation}
The ratio of the merger time scale to the Hubble time-scale is thus
\begin{equation}
    \frac{\tau_{\rm merger}}{1/H}\sim  10^{28}\frac{M_P^6}{T^4M^3}\left(\frac{M_P}{M}\right)^{-4/3}\frac{T^2}{M_P}\sim 10^{28}\left(\frac{M_P}{T}\right)^2\left(\frac{M_P}{M}\right)^{5/3}\gg 1,
\end{equation}
thus mergers are completely irrelevant to shaping the mass functions of DBBHs.


\section{Conclusions}

This study considered the possibility that SU(N) dark-sector theories produce Planck-scale black holes upon the collapse of dark baryons. 
We showed that DBBHs can naturally arise  if either the confinement scale or the quark mass is close to the Planck scale, assuming dark mesons and glueballs decay to visible-sector particles. In the large $N$ limit, the dark baryon abundance is exponentially suppressed with $N$, but it receives a significant contribution from freeze-in processes where two mesons or glueballs produce out-of-equilibrium baryon-antibaryon pairs.

DBBHs can have the right abundance to be the dark matter -- provided Hawking evaporation is suppressed -- for several combinations of $N$, the dark-to-visible sector temperature ratios, and the effective exponential suppression for the production cross section of dark baryons. An alternate possibility, arising for sufficiently hot dark sectors and low $N$, is that the DBBH abundance is large enough that it could produce the right dark matter abundance even upon inflationary dilution.

Because of the exponential suppression with $N$, we find that even for very hot dark sectors, $N$ cannot exceed values around 100, implying that the mass of the produced DBBHs cannot exceed a few hundreds times the Planck mass. This implies DBBH masses at most of a few milligrams.

Finally, the requirement that the confinement phase transition occur at temperatures below  reheating forces hot dark sectors in the light-quark regime, barring exotic inflationary scenarios that allow Planck-scale reheating temperatures, but it is easily satisfied for Planck-scale quark masses in the heavy quark regime.

We showed that the resulting DBBH population is entirely unaffected by either accretion or mergers, and it produces an effect on the matter power spectrum only at extremely small scales.

Planck-scale relics can potentially acquire electric charge, which would make them detectable, most prominently with so-called paleo-detectors \cite{Lehmann:2019zgt, Profumo:2024fxq}. Additionally, mergers of DBBHs may lead to both the production of high-frequency gravitational waves, and to the production of particles if the resulting final mass is above the threshold for evaporation. 

In the Appendix we discuss how our results are effected taking into account the possible renormalization group evolution of Newton's gravitational constant. We find that the overall picture is unchanged, but that the requirement of collapse into DBBH presents stricter constraints, limiting the maximal DBBH mass to roughly 6 units of the bare Planck scale.

As a final remark, we note that if dark baryons are sub-critical, i.e. have a size close, but smaller, than their Schwarzschild radius, it is possible that bound ``dark nuclei'' form, mediated by dark glueballs and mesons. Such objects may, in principle, and in turn, collapse into ``dark nuclei black holes''. We leave this possibility for future study, although naive estimates based on the scaling of the dark nuclei size and mass with the dark nucleon number $A$,
\begin{equation}
R_N\sim \frac{A^{1/3}}{\Lambda},\qquad M_N\sim A\ {\rm max}\left(\Lambda,m_Q\right)
\end{equation}
suggest that dark nuclei black hole would form for sufficiently large $A$
\begin{equation}
    A\gg\left(\frac{M_P}{\Lambda}\right)^3\ (\Lambda>m_Q);\qquad A\gg\left(\frac{M^2_P}{\Lambda m_Q}\right)^{3/2}\ (\Lambda<m_Q).
\end{equation}
Large $A\sim 10^6-10^9$ are arguably possible, for extremely favorable confinement dynamics, see e.g. \cite{Krnjaic:2014xza, Redi:2018muu}.

\section*{Acknowledgements} \label{sec:acknowledgements}
   \noindent  The Author is grateful to Alexander Profumo for inspiring the original idea that prompted this study, and to Aarna Garg, Sefi Katznelson, Wolfgang Altmannshofer, Stefania Gori, Edgar Shaghoulian and Rouven Balkin for useful remarks, thoughts, and feedback. The Author is also especially thankful to Tim Cohen, Alexander Kusenko, and Graham Kribbs for crucial feedback on this manuscript. This material is based upon work supported in part by the U.S. Department of Energy grant number de-sc0010107.

\appendix
\section{Possible Effects of the Renormalization Group Evolution of Newton's Gravitational Constant}\label{app:RGMP}

\begin{figure}
    \centering
    \includegraphics[width=0.5\linewidth]{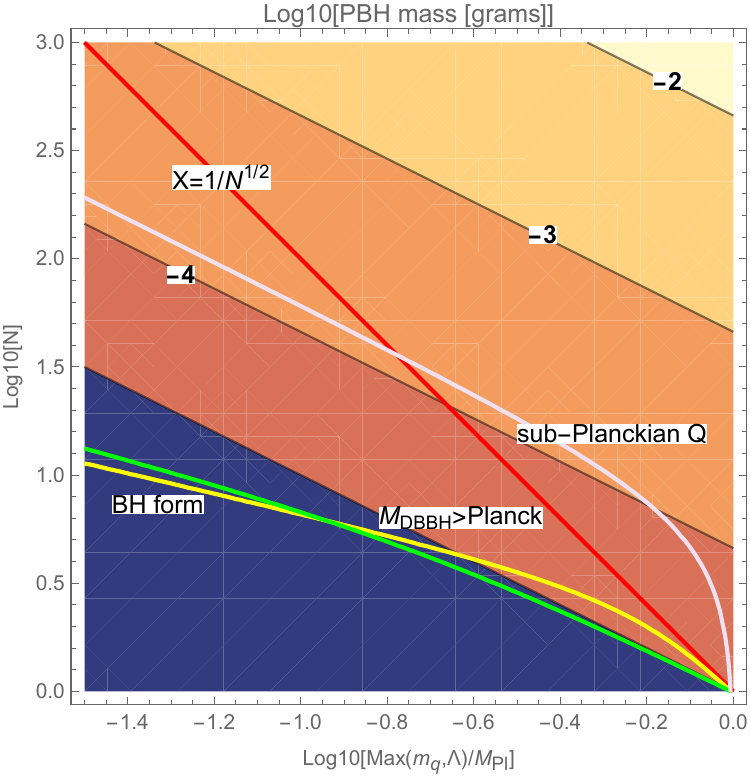}
         \caption{The ``Planck-scale-renormalized'' results; see the text for details.}
    \label{fig:RGcorr}
\end{figure}

Assuming that the running of Newton's Gravitational Constant, and hence of the Planck mass, follows the results of Ref.~\cite{Calmet:2008tn}, we postulate the quadratic running
\begin{equation}
M_P^2(\mu)=M_P^2-\frac{N_{\rm dof}\mu^2}{12\pi}.
\end{equation}
Neglecting the possibility of a large number of flavors $N_f$ in the large $N$ limit, i.e. for $N^2\gg N_f$, and further neglecting possible additional dark scalars (that would contribute with the opposite sign to the running \cite{Calmet:2008tn}) the {\em RG-corrected} condition that black holes form from dark baryons reads\footnote{For simplicity, we use here $m_Q$ instead of min$(\Lambda,m_Q)$.}
\begin{equation}
    \frac{1}{m_Q}<\frac{Nm_Q}{M_P^2-\frac{N^2(Nm_Q)^2}{12\pi}},
\end{equation}
where we indicate with $M_P$ the ``bare'' Planck mass. This condition then reads, in terms of $X=m_Q/M_P$,
\begin{equation}
    X>\sqrt{\frac{1}{N+N^{13/3}/(12\pi)}}.
\end{equation}
The condition that $M_{\rm DBBH}>M_P(M_{\rm DBBH})$ reads instead
\begin{equation}
    X>\left(N^2+N^4/(12\pi)\right)^{-1/2}.
\end{equation}
Finally, the condition that $m_Q<M_P(m_Q)$ reads
\begin{equation}
    X<\left(1+N^2/(12\pi)\right)^{-1/2}.
\end{equation}
Taking the asymptotic large $N$ behaviors of the expressions above, we get that
\begin{equation}
    \frac{1}{N^{2}}\lesssim X\lesssim\frac{1}{N}.
\end{equation}
We show the corresponding numerical results in fig.~\ref{fig:RGcorr}; the allowed region is between the grey line, corresponding to the requirement that $m_Q$ be sub-Planckian, and the yellow line, corresponding to the formation condition. The latter is more stringent than the condition that the black hole mass be trans-Planckian. The plot also highlgihts the asymptotic maximal DBBH mass.

We note that the application of renormalization group (RG) techniques to evolve the Planck scale, while intriguing, remains highly speculative and fraught with uncertainties. Perturbative methods  employed in RG calculations may break down at the Planck scale due to the non-renormalizability of gravity \cite{Goroff:1985th} and the potential onset of strong coupling effects \cite{Donoghue:1994dn}. The assumption of a fixed background spacetime in many RG approaches \cite{Reuter:1996cp} may also be inappropriate at scales where quantum gravitational effects become significant. Furthermore, the possible existence of additional new physics, as noted above, or fundamental discreteness at the Planck scale \cite{Rovelli:1994ge} could invalidate the continuous RG flow picture. 

\bibliographystyle{apsrev4-1}
\bibliography{bib}
\end{document}